\documentclass[sigconf,natbib=true]{acmart}

\usepackage{balance} 
\usepackage{enumitem}

\usepackage{pifont}

\newcommand{\blackcircled}[1]{\ding{\numexpr201+#1\relax}}

\AtBeginDocument{%
  }

\setcopyright{acmlicensed}

\copyrightyear{2026}
\acmYear{2026}
\setcopyright{cc}
\setcctype{by}
\acmConference[CHIIR '26]{2026 ACM SIGIR Conference on Human Information Interaction and Retrieval}{March 22--26, 2026}{Seattle, WA, USA}
\acmBooktitle{2026 ACM SIGIR Conference on Human Information Interaction and Retrieval (CHIIR '26), March 22--26, 2026, Seattle, WA, USA}
\acmPrice{}
\acmDOI{10.1145/3786304.3787914}
\acmISBN{979-8-4007-2414-5/2026/03}

\begin{document}

\title{Seek and You Shall Find: Design \& Evaluation of a Context-Aware Interactive Search Companion}

\author{Markus Bink}
\affiliation{
  \institution{Neu-Ulm University of Applied Sciences}
  \city{Neu-Ulm}
  \country{Germany}}
\affiliation{
  \institution{University of Regensburg}
  \city{Regensburg}
  \country{Germany}  
  }
\email{markus.bink@ur.de}

\author{Marten Risius}
\affiliation{
  \institution{Neu-Ulm University of Applied Sciences}
  \city{Neu-Ulm}
  \country{Germany}}
\email{marten.risius@hnu.de}

\author{Udo Kruschwitz}
\affiliation{
  \institution{University of Regensburg}
  \city{Regensburg}
  \country{Germany}}
\email{udo.kruschwitz@ur.de}

\author{David Elsweiler}
\affiliation{
  \institution{University of Regensburg}
  \city{Regensburg}
  \country{Germany}}
\email{david.elsweiler@ur.de}

\renewcommand{\shortauthors}{Bink et al.}

\begin{abstract}
Many users struggle with effective online search and critical evaluation, especially in high-stakes domains like health, while often overestimating their digital literacy. Thus, in this demo, we present an interactive search companion that seamlessly integrates expert search strategies into existing search engine result pages. Providing context-aware tips on clarifying information needs, improving query formulation, encouraging result exploration, and mitigating biases, our companion aims to foster reflective search behaviour while minimising cognitive burden. A user study demonstrates the companion's successful encouragement of more active and exploratory search, leading users to submit 75\% more queries and view roughly twice as many results, as well as performance gains in difficult tasks. This demo illustrates how lightweight, contextual guidance can enhance search literacy and empower users through micro-learning opportunities. While the vision involves real-time LLM adaptivity, this study utilises a controlled implementation to test the underlying intervention strategies.
\end{abstract}

\ccsdesc[500]{Human-centered computing~User studies}
\ccsdesc[500]{Information systems~Search interfaces}

\keywords{search behaviour, companion, nudging, boosting, learning to search}

\begin{teaserfigure}
  \frame{\includegraphics[width=\textwidth]{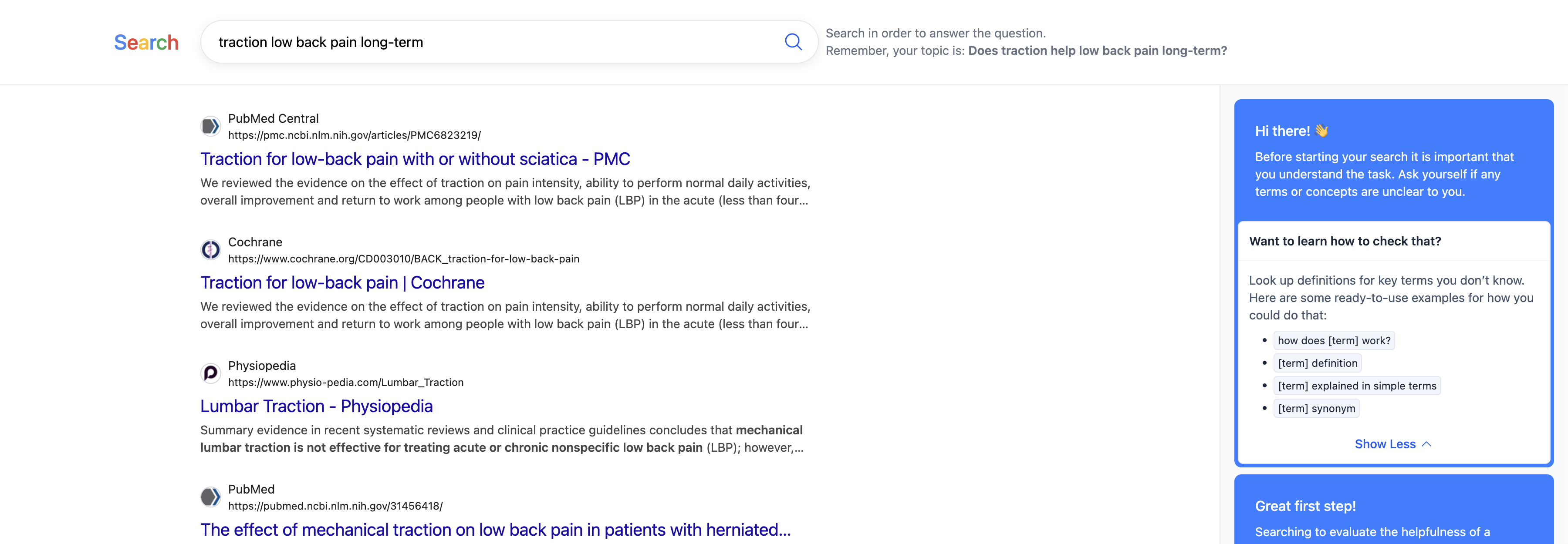}}
  \caption{Example SERP with search results on the left and the proposed interactive search companion on the right}
  \label{fig:serp}
\end{teaserfigure}

\maketitle

\section{Introduction \& Related Work}
Despite unprecedented access to online information, many people still struggle to search effectively and evaluate sources critically. In high-stakes domains such as health, limited search literacy can lead to misinterpretation of results and poor decision-making \cite{koopman2023dr, butler2019health, ghenai2017health, do2022infodemics, fox2013health, eurostat_digital_economy_households}. This effect is compounded by many users overestimating their online information literacy skills \cite{mahmood2016people}.

Prior research has explored pedagogically inspired, chat-based search copilots that prompt reflection and aim to improve search literacy \cite{Bink2026CanYouTellMe}. However, such systems have been found to increase cognitive load and suffer from usability issues when users must invest substantial effort to engage with instructional content during active search \cite{Bink2026CanYouTellMe}. This burden can frustrate users and limit adoption, offsetting potential learning benefits and hindering search outcomes. This inspired us to develop a lightweight, seamlessly integrated search UI feature that encourages reflective searching without disrupting the user’s flow.

One way to conceptualise such interventions is through the lens of \emph{nudging} \cite{thaler2009nudge} and \emph{boosting} \cite{hertwig2017nudging}. Nudge-based strategies, such as augmenting search results with credibility cues or privacy indicators, have been shown to influence search behaviour \cite{zimmerman2019privacy, yamamoto2011enhancing, schwarz2011augmenting}, but are often viewed as paternalistic \cite{hertwig2017nudging} and their effects tend to diminish once the intervention is removed \cite{hertwig2017nudging, ortloff2021effect}. Boosts, in contrast, aim to empower users by providing competences with the potential to persist beyond the intervention \cite{hertwig2017nudging}. For example, showing users how their performance compares to experts can influence interaction patterns \cite{bateman2012search}, and brief search tips can positively shape behaviour on SERPs \cite{bink2024balancing}.

While such interventions show promise, most remain permanently visible offering guidance regardless of whether users currently need it. Building on these insights, we present an \emph{interactive search companion} demo\footnote{\url{https://demo-interactive-search-companion.vercel.app/}} that illustrates how context-aware guidance can scaffold expert search behaviour — clarifying intent, formulating effective queries, and evaluating results — directly into the search interface. By providing brief, context-specific tips at key stages of the search process, the companion aims to support users when and where guidance is needed while minimising cognitive load. The demo illustrates this concept through a controlled, static prototype. We chose this implementation to ensure all participants received identical guidance, allowing us to isolate the effects of the intervention strategies without the potential variability of real-time LLM generation.
\begin{figure*}[!tbp]
  \frame{\includegraphics[width=\textwidth]{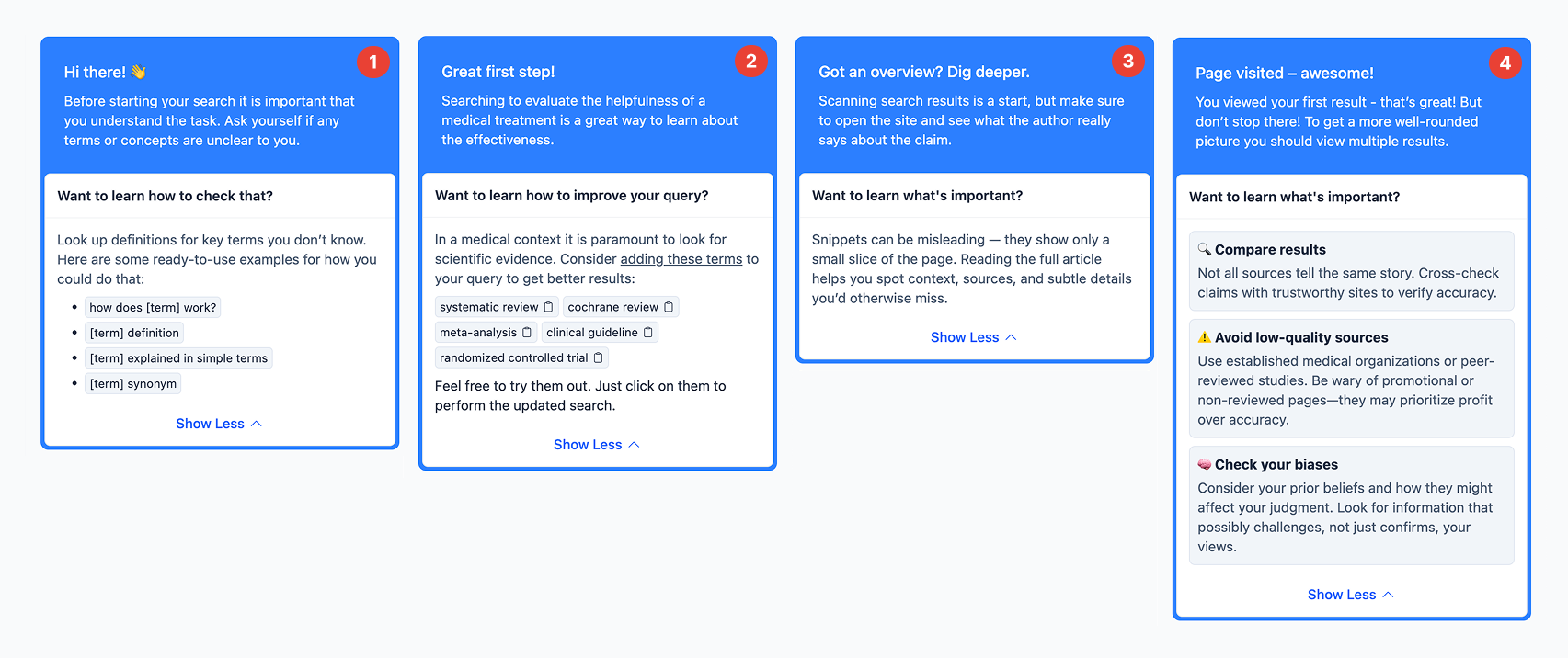}}
  \caption{Overview of all interactive search tips. For illustration purposes, these are presented in their expanded form to show all additional content if users click on 'Show More'. (\textit{Numbers in red are not part of the companion but used for referencing)}}
  \label{fig:overview-search-tips}
\end{figure*}

\section{Implementation}\label{sec:implementation}
The following section describes the interactive search companion interface, outlining each search tip’s components and how they are presented based on users’ contextual interactions with the SERP. 

\textbf{Interactive Search Companion.} Designed for integration into commercial SERPs (e.g., Google, Bing), the companion appears as a right-hand sidebar (see Figure \ref{fig:serp}). To visually distinguish it from the SERP, the companion has a subtle grey background on which contextual search tips are displayed. The sidebar functions as a scrollable container, allowing users to revisit earlier tips at any time. Search tips are presented in chronological order, with the most recent appearing at the bottom. When a new tip is introduced, the sidebar automatically scrolls to bring it fully into view, ensuring that users remain aware of new guidance while maintaining the flow of their search activity.

The companion is designed based on expert search behaviour and information literacy principles. For instance, it encourages users to view and compare multiple results \cite{bateman2012search, white2008medical}, thereby supporting lateral reading strategies \cite{wineburg2017lateral, panizza2022lateral}. 
Additionally, it teaches domain-specific query generation principles, illustrated through examples such as searching for systematic reviews in health contexts (e.g., “Cochrane review”), to promote higher-quality results. It achieves context-awareness by monitoring search states (e.g., dwell time, query submission) to trigger interventions, and provides interactivity through actionable UI elements that reduce the effort of adopting new strategies.

\textbf{Search Tip Overview.} Each tip operationalises a distinct phase of expert search behaviour, combining micro-learning design with low-friction interaction (see Figure \ref{fig:overview-search-tips} for an overview): 

\underline{Headline:} A concise title designed to capture users’ attention and convey the essence of the tip. 

\underline{Description:} A brief explanatory text that offers a high-level recommendation tailored to the user’s current stage in the search process.

\underline{Learning Opportunity:} For users interested in improving their search literacy, an expandable accordion is placed at the bottom of each tip. Its title is phrased as a question, and the accordion remains collapsed by default, revealing only a teaser of the additional content. This design intentionally leverages users’ curiosity to encourage deeper engagement.

While some users may already be familiar with the presented strategies, the information remains valuable to both novice and expert searchers. Novices can discover new strategies and query formulation techniques, while experts receive a subtle refresher that helps bring prior knowledge to the forefront. 

The content is action-oriented, providing concrete, clickable suggestions that users can apply directly within their ongoing search session. This approach reduces friction by delivering relevant information at key stages of the search process, while the interactive elements allow users to apply strategies seamlessly, thus decreasing cognitive effort.

To maximise experimental control, the tips in this demo were pre-defined, thereby eliminating the variability inherent in real-time generation.

\textbf{\blackcircled{1} Clarifying Information Need.}
The first search tip supports users’ sense-making by helping them clarify key terms and concepts relevant to their task. In a medical search scenario, for instance, users may encounter unfamiliar terminology that hinders understanding. To address this, the system presents ready-to-use, actionable queries that users can employ to clarify unknown terms before proceeding.

\textbf{\blackcircled{2} Improving Query Reformulation.}
The second tip focuses on enhancing users’ query literacy, particularly within medical health searches. It introduces effective reformulation strategies, such as adding modifiers like systematic review to retrieve more trustworthy sources. Suggestions are clickable, automatically submitting the revised query to the search engine. This design reduces user effort and offers an interactive, low-friction way to observe how different formulations affect search outcomes – supporting quick, experiential learning.

\textbf{\blackcircled{3} Result Exploration.} The third search tip is intended to encourage exploration by viewing individual search results rather than just skimming the SERP. In many cases, the search result snippets often already contain an answer to users' questions \cite{white2013beliefs}, however, the extracted snippet might not tell the whole picture the linked website is trying to tell. Therefore, users are encouraged to visit webpages.

\textbf{\blackcircled{4} Bias Mitigation and Quality Assurance.}
The final tip aims to strengthen critical evaluation and bias awareness. Viewing one result is a start, but gaining a well-rounded understanding requires consulting multiple sources. The tip introduces lateral reading strategies \cite{panizza2022lateral, wineburg2017lateral}—encouraging users to compare results and cross-check claims—and reminds them to prioritize trustworthy medical sources. It also highlights common cognitive pitfalls such as confirmation bias \cite{azzopardi2021cognitive}, which can influence result selection.

\textbf{Triggering Search Tips.} The first search tip \blackcircled{1} is shown when users initially access the search engine. It aims to raise awareness of potential gaps in their understanding of the search task, such as unfamiliar terms or concepts, to ensure that relevant terminology is correctly interpreted and can support accurate conclusions in later stages.

The second search tip \blackcircled{2} appears after users submit their first query. Given the medical context, it provides appropriate query suggestions that users can directly adopt for further searches.

The third search tip \blackcircled{3} is optional and not displayed to all users. Prior work shows that search result snippets often contain answers to users’ questions \cite{white2013beliefs, Bink2026CanYouTellMe}, leading many to rely solely on these when forming opinions. In the medical domain, however, accessing the full context is critical, as snippets may present an incomplete or misleading picture. Accordingly, if users do not interact with any search result within 20 seconds after submitting their initial query, this tip appears to explain the importance of engaging with the underlying content. The 20-second threshold is derived from previously collected interaction data \cite{Bink2026CanYouTellMe}. To avoid overloading users or interrupting fluent exploration — a key concern in designing pedagogical copilots — the system introduces this tip only if no result interaction occurs within 20 seconds.

The fourth and final search tip \blackcircled{4} is displayed when users return to the results page after visiting a document for the first time.

\begin{table*}[t]
\centering
\caption{Search outcome accuracy across condition and topic in percent.}
\label{tab:search_outcome_accuracy}
\begin{tabular}{lllllll}
\toprule
              & \textsc{antioxidants}  & \textsc{benzodiazepines} & \textsc{caffeine}     & \textsc{melatonin}    & \textsc{probiotics}    & \textsc{traction}    \\
\midrule
\textsc{10-blue-links} & 71.4          & \textbf{100}    & \textbf{100} & \textbf{100} & 38.9          & 57.1        \\
\textsc{companion}     & \textbf{76.9} & 90              & 90.9         & 92.3         & \textbf{41.2} & \textbf{60} \\
\bottomrule
\end{tabular}
\end{table*}
\section{Evaluation}\label{sec:evaluation}
To evaluate the interactive search companion, we conducted a pre-registered\footnote{The pre-registration containing full methodological details can be found here: \url{https://osf.io/rnpzv/overview?view_only=1c7f9fb3d86f4fbc9a68c7f44308a215}} between-groups user study comparing two search systems. In the baseline condition, users were presented with the standard \textsc{10-blue-links}, whereas in the experimental condition, the \textsc{interactive search companion} provided contextual search tips at key points during the search process. Participants were randomly assigned to one of these experimental conditions and a single  medical search task (e.g., \textit{"Do probiotics help treat eczema?"})~\cite{pogacar2017positive} and had to determine, based on their search, whether the treatment was helpful. We selected the medical domain because people frequently consult search engines for health-related issues \cite{fox2013health}, but the quality of information varies across sites \cite{daraz2019can}, highlighting the need for user support in this critical domain. In addition to task outcomes, we recorded user interactions with search results and queries issued, evaluating whether the companion encourages reflective and exploratory search, focusing on answer accuracy (H1), result exploration (H2), and query formulation (H3). We also tracked engagement with the search tips, capturing which tips were presented and how users interacted with them. Based on these data, we formulated the following hypotheses:

\begin{itemize}[leftmargin=*]
    \item \textbf{H1:}  \textit{Users assisted by the search companion are more likely to correctly answer a medical question.}  
    
    \item \textbf{H2:} \textit{Users assisted by the search companion will view more search results.}
    
    \item \textbf{H3:} \textit{Users assisted by the search companion will issue more queries.} 
\end{itemize}

 170 participants, fluent in English to ensure full comprehension of presented information, were recruited via the Prolific platform. The sample size was determined through an a priori power analysis in G*Power using a Chi-Square test with an effect size of 0.25, a power of 0.8, and a significance level ($\alpha$) of 0.0167. The reduced $\alpha$ results from a Bonferroni correction applied for testing three hypotheses ($0.05 / 3 = 0.0167$).

\section{Results}

Overall search accuracy was similar for the baseline (73.2 \%) and the companion-assisted system (73.0 \%). However, task-level analysis reveals topic-related differences (see Table \ref{tab:search_outcome_accuracy}). For more difficult tasks—indicated by lower accuracies in the 10-blue-links condition—the companion’s guidance slightly improved outcomes, suggesting its benefits may depend on task difficulty, a point revisited in the discussion.
On average, the search companion ($M = 2.64, SD = 1.52$) influenced users to view about twice as many search results compared to the baseline ($M = 1.30, SD = 1.10$, $p < .001$). A similar trend was found for query formulations with users supported by the companion issuing more queries ($M = 1.96, SD = 1.21$) than those without ($M = 1.12, SD = 0.41$, $p < .001$).


\begin{table}[t]
\centering
\caption{Frequencies of search tip presentation and user engagement with "Learn More" component}
\label{tab:search_tips_frequency}
\begin{tabular}{lcc}
\toprule
\textbf{Search Tip} & \textbf{Shown} & \textbf{Opened} \\
\midrule
Clarify information need \blackcircled{1} & 74/74 & 48/74 \\
Optimize query \blackcircled{2} & 73/74 & 43/73 \\
Result exploration \blackcircled{3} & 47/74 & 23/47 \\
Compare results \blackcircled{4} & 69/74 & 46/69 \\
\bottomrule
\end{tabular}
\end{table}

Participants interacted frequently with the search tips presented by the system (see Table \ref{tab:search_tips_frequency}). Among those who explored the optimize query tip \blackcircled{2}, 17 of 43 users interacted with the embedded query suggestions. Most (11/17) selected a single suggestion; others tried up to four. Notably, investigating the queries revealed that several users who did not click on any suggestions nonetheless incorporated similar modifications into their subsequent queries, suggesting strategy uptake even without direct interaction. This is further supported by users feedback mentioning that the companion "\textit{[...] gave me better support terms in my search}",  "\textit{[...] it encouraged me to look at more articles to balance my view}" and that they \textit{"[...] found it quite helpful}" and "\textit{[...] something I would use}". 

\section{Discussion}

We observed significant positive differences in user behaviour: participants with the companion submitted more queries and viewed more search results. While overall accuracy did not differ significantly across all tasks, participants in the companion condition showed gains on all but the easiest ones, where performance was already at the ceiling in the 10-blue-links condition.

Users typically click only a few top-ranked results \cite{bateman2012search, joachims2007evaluating}. In contrast, a single intervention with our companion doubled the number of pages viewed, outperforming earlier boost-style approaches that failed to increase engagement \cite{bink2024balancing, Bink2026CanYouTellMe}. Users also reformulated queries more often and adopted domain-relevant terms, indicating more advanced strategies \cite{aula2003query, white2009characterizing,elsweiler2025query}.

Overall, the companion encouraged active, exploratory behaviour, helpful for complex tasks but less so for simpler ones, where interventions slightly impaired performance. Additional guidance may have triggered unnecessary reformulations or diversions from correct answers, possibly due to cognitive overload or overthinking, causing users to second-guess their initial assessment.

These results highlight a trade-off between supporting immediate task performance, fostering long-term learning, and risking performance loss when focus shifts from doing to reflecting. Balancing assistance and autonomy thus remains a key design challenge for search companions. Further analysis of user interactions will help clarify whether performance differences stem from cognitive factors, result positioning, or interface design.

\section{Conclusion \& Future Work}

In summary, the presented demo shows how an interactive search companion can effectively influence users’ search behaviour, encouraging more active, exploratory engagement and fostering micro-learning opportunities such as query formulation. Although this behavioural shift did not consistently enhance answer correctness across all tasks, it proved particularly valuable in more challenging ones. These preliminary results highlight the potential of interactive companions to strengthen user agency and meta-cognitive awareness during search. Moving forward, we aim to analyse variations in user interaction and cognitive biases that may have affected result selection and limited performance gains. Based on these insights, we will refine the companion and test it in more complex and contested search settings, introducing additional tips and triggers. We also plan to examine how the companion can complement GenAI systems that provide direct answers, and to develop personalised controls allowing users to tailor the level and type of support they receive. Ultimately, we envision the companion as a browser extension, enabling users to cultivate effective search strategies through everyday use. At the demo, attendees can explore the companion’s contextual guidance in real time, observing how micro-interventions shape their own search patterns.

\begin{acks}
This work receives generous funding support from the Bavarian State Ministry of Science and the Arts through the Distinguished Professorship Program as part of the Bavarian High-Tech Agenda.
\end{acks}

\bibliographystyle{ACM-Reference-Format}
\bibliography{bibliography}

@inproceedings{elsweiler2025query,
  title={Query Smarter, Trust Better? Exploring Search Behaviours for Verifying News Accuracy},
  author={Elsweiler, David and Ateia, Samy and Bink, Markus and Donabauer, Gregor and Fern{\'a}ndez Pichel, Marcos and Frummet, Alexander and Kruschwitz, Udo and Losada, David E and Ludwig, Bernd and Meyer, Selina and others},
  booktitle={Proceedings of the 48th International ACM SIGIR Conference on Research and Development in Information Retrieval},
  pages={515--526},
  year={2025}
}

@inproceedings{ghenai2017health,
  title={Health misinformation in search and social media},
  author={Ghenai, Amira},
  booktitle={Proceedings of the 2017 International Conference on Digital Health},
  pages={235--236},
  year={2017}
}

@article{do2022infodemics,
  title={Infodemics and health misinformation: a systematic review of reviews},
  author={Do Nascimento, Israel Junior Borges and Pizarro, Ana Beatriz and Almeida, Jussara M and Azzopardi-Muscat, Natasha and Gon{\c{c}}alves, Marcos Andr{\'e} and Bj{\"o}rklund, Maria and Novillo-Ortiz, David},
  journal={Bulletin of the World Health Organization},
  volume={100},
  number={9},
  pages={544},
  year={2022}
}

@inproceedings{zimmerman2019privacy,
  title={Privacy nudging in search: Investigating potential impacts},
  author={Zimmerman, Steven and Thorpe, Alistair and Fox, Chris and Kruschwitz, Udo},
  booktitle={Proceedings of the 2019 Conference on Human Information Interaction and Retrieval},
  pages={283--287},
  year={2019}
}

@article{mahmood2016people,
  title={Do people overestimate their information literacy skills? A systematic review of empirical evidence on the Dunning-Kruger effect},
  author={Mahmood, Khalid},
  journal={Communications in Information Literacy},
  volume={10},
  number={2},
  pages={3},
  year={2016}
}

@article{wineburg2017lateral,
  title={Lateral reading: Reading less and learning more when evaluating digital information},
  author={Wineburg, Sam and McGrew, Sarah},
  year={2017},
  publisher={Stanford history education group working paper}
}

@inproceedings{bink2024balancing,
  title={Balancing Act: Boosting Strategies for Informed Search on Controversial Topics},
  author={Bink, Markus and Elsweiler, David},
  booktitle={Proceedings of the 2024 Conference on Human Information Interaction and Retrieval},
  pages={254--265},
  year={2024}
}

@article{daraz2019can,
  title={Can Patients Trust Online Health Information? A Meta-narrative Systematic Review Addressing the Quality of Health Information on the Internet},
  author={Daraz, Lubna and Morrow, Allison S and Ponce, Oscar J and Beuschel, Bradley and Farah, Magdoleen H and Katabi, Abdulrahman and Alsawas, Mouaz and Majzoub, Abdul M and Benkhadra, Raed and Seisa, Mohamed O and others},
  journal={Journal of general internal medicine},
  volume={34},
  pages={1884--1891},
  year={2019},
  publisher={Springer}
}

@article{fox2013health,
  title={Health Online 2013},
  author={Fox, Susannah and Duggan, Maeve},
  journal={Health},
  volume={2013},
  pages={1--55},
  year={2013}
}

@inproceedings{white2013beliefs,
  title={Beliefs and Biases in Web Search},
  author={White, Ryen},
  booktitle={Proceedings of the 36th international ACM SIGIR conference on Research and development in information retrieval},
  pages={3--12},
  year={2013}
}

@inproceedings{koopman2023dr,
  title={Dr ChatGPT tell me what I want to hear: How different prompts impact health answer correctness},
  author={Koopman, Bevan and Zuccon, Guido},
  booktitle={Proceedings of the 2023 conference on empirical methods in natural language processing},
  pages={15012--15022},
  year={2023}
}

@inproceedings{schwarz2011augmenting,
author = {Schwarz, Julia and Morris, Meredith},
title = {Augmenting Web Pages and Search Results to Support Credibility Assessment},
year = {2011},
isbn = {9781450302289},
publisher = {Association for Computing Machinery},
address = {New York, NY, USA},
url = {https://doi.org/10.1145/1978942.1979127},
doi = {10.1145/1978942.1979127},
booktitle = {Proceedings of the SIGCHI Conference on Human Factors in Computing Systems},
pages = {1245–1254},
numpages = {10},
location = {Vancouver, BC, Canada},
series = {CHI '11}
}

@inproceedings{yamamoto2011enhancing,
author = {Yamamoto, Yusuke and Tanaka, Katsumi},
title = {Enhancing Credibility Judgment of Web Search Results},
year = {2011},
isbn = {9781450302289},
publisher = {Association for Computing Machinery},
address = {New York, NY, USA},
url = {https://doi.org/10.1145/1978942.1979126},
doi = {10.1145/1978942.1979126},
booktitle = {Proceedings of the SIGCHI Conference on Human Factors in Computing Systems},
pages = {1235–1244},
numpages = {10},
location = {Vancouver, BC, Canada},
series = {CHI '11}
}

@article{joachims2007evaluating,
author = {Joachims, Thorsten and Granka, Laura and Pan, Bing and Hembrooke, Helene and Radlinski, Filip and Gay, Geri},
title = {Evaluating the Accuracy of Implicit Feedback from Clicks and Query Reformulations in Web Search},
year = {2007},
issue_date = {April 2007},
publisher = {Association for Computing Machinery},
address = {New York, NY, USA},
volume = {25},
number = {2},
issn = {1046-8188},
url = {https://doi.org/10.1145/1229179.1229181},
doi = {10.1145/1229179.1229181},
journal = {ACM Trans. Inf. Syst.},
month = {apr},
pages = {7–es},
numpages = {27}
}

@inproceedings{pogacar2017positive,
author = {Pogacar, Frances A. and Ghenai, Amira and Smucker, Mark D. and Clarke, Charles L.A.},
title = {The Positive and Negative Influence of Search Results on People's Decisions about the Efficacy of Medical Treatments},
year = {2017},
isbn = {9781450344906},
publisher = {Association for Computing Machinery},
address = {New York, NY, USA},
url = {https://doi.org/10.1145/3121050.3121074},
doi = {10.1145/3121050.3121074},
booktitle = {Proceedings of the ACM SIGIR International Conference on Theory of Information Retrieval},
pages = {209–216},
numpages = {8},
location = {Amsterdam, The Netherlands},
series = {ICTIR '17}
}

@book{thaler2009nudge,
  title={Nudge: Improving decisions about health, wealth, and happiness},
  author={Thaler, Richard H and Sunstein, Cass R},
  year={2009},
  publisher={Penguin}
}

@inproceedings{ortloff2021effect,
author = {Ortloff, Anna-Marie and Zimmerman, Steven and Elsweiler, David and Henze, Niels},
title = {The Effect of Nudges and Boosts on Browsing Privacy in a Naturalistic Environment},
year = {2021},
isbn = {9781450380553},
publisher = {Association for Computing Machinery},
address = {New York, NY, USA},
url = {https://doi.org/10.1145/3406522.3446014},
doi = {10.1145/3406522.3446014},
booktitle = {Proceedings of the 2021 Conference on Human Information Interaction and Retrieval},
pages = {63–73},
numpages = {11},
location = {Canberra ACT, Australia},
series = {CHIIR '21}
}

@article{hertwig2017nudging,
  title={Nudging and Boosting: Steering or Empowering Good Decisions},
  author={Hertwig, Ralph and Gr{\"u}ne-Yanoff, Till},
  journal={Perspectives on Psychological Science},
  volume={12},
  number={6},
  pages={973--986},
  year={2017},
  publisher={Sage Publications Sage CA: Los Angeles, CA}
}

@article{panizza2022lateral,
  title={Lateral reading and monetary incentives to spot disinformation about science},
  author={Panizza, Folco and Ronzani, Piero and Martini, Carlo and Mattavelli, Simone and Morisseau, Tiffany and Motterlini, Matteo},
  journal={Scientific Reports},
  volume={12},
  number={1},
  pages={5678},
  year={2022},
  publisher={Nature Publishing Group UK London}
}

@article{butler2019health,
  title={Health information seeking behaviour: the librarian's role in supporting digital and health literacy},
  author={Butler, Rachel},
  journal={Health Information \& Libraries Journal},
  volume={36},
  number={3},
  pages={278--282},
  year={2019},
  publisher={Wiley Online Library}
}

@misc{eurostat_digital_economy_households,
  author       = {Eurostat},
  title        = {Digital economy and society statistics – households and individuals},
  howpublished = {\url{https://ec.europa.eu/eurostat/statistics-explained/index.php?title=Digital_economy_and_society_statistics_-_households_and_individuals}},
  year         = {2025}, 
  note         = {last accessed October 2025}
}

@inproceedings{bateman2012search,
  title={The search dashboard: how reflection and comparison impact search behavior},
  author={Bateman, Scott and Teevan, Jaime and White, Ryen W},
  booktitle={Proceedings of the SIGCHI Conference on Human Factors in Computing Systems},
  pages={1785--1794},
  year={2012}
}

@inproceedings{azzopardi2021cognitive,
  title={Cognitive biases in search: a review and reflection of cognitive biases in Information Retrieval},
  author={Azzopardi, Leif},
  booktitle={Proceedings of the 2021 conference on human information interaction and retrieval},
  pages={27--37},
  year={2021}
}

@inproceedings{white2008medical,
  title={How medical expertise influences web search interaction},
  author={White, Ryen W and Dumais, Susan and Teevan, Jaime},
  booktitle={Proceedings of the 31st annual international ACM SIGIR conference on Research and development in information retrieval},
  pages={791--792},
  year={2008}
}

@inproceedings{white2009characterizing,
  title={Characterizing the influence of domain expertise on web search behavior},
  author={White, Ryen W and Dumais, Susan T and Teevan, Jaime},
  booktitle={Proceedings of the second ACM international conference on web search and data mining},
  pages={132--141},
  year={2009}
}

@inproceedings{aula2003query,
  title={Query Formulation in Web Information Search.},
  author={Aula, Anne},
  booktitle={ICWI},
  pages={403--410},
  year={2003}
}

@inproceedings{Bink2026CanYouTellMe,
  title     = {{``Can You Tell Me?''}: Designing Copilots to Support Human Judgement in Online Information Seeking},
author    = {Bink, Markus and Risius, Marten and Kruschwitz, Udo and Elsweiler, David},
  booktitle = {Proceedings of the 2026 Conference on Human Information Interaction and Retrieval},
  year      = {2026},
  doi       = {10.1145/3786304.3787866},
}

\end{document}